\documentclass[prb, letterpaper, showpacs,twocolumn, floatfix, superscriptaddress, balancelastpage, 10pt]{revtex4}
\usepackage{graphicx, dcolumn,graphicx,color,booktabs}
\usepackage{microtype}
\usepackage[colorlinks,plainpages=false,linkcolor=black,urlcolor=blue,citecolor=blue,pdfpagemode=UseNone,pdfstartview=FitH]{hyperref}

\newcommand \sample{Sr$_{\scriptsize{\textup{2}}}$RuO$_{\scriptsize{\textup{4}}}${}}

\renewcommand{\figurename}{Fig.}
\renewcommand{\tablename}{Table}
\makeatletter\renewcommand{\fnum@figure}[1]{\figurename~\thefigure}\makeatother
\makeatletter\renewcommand{\fnum@table}[1]{\tablename~\thetable.}\makeatother

\begin{document}
\title{Surface and bulk electronic structure of unconventional superconductor Sr$_2$RuO$_4$: unusual splitting of the $\beta$ band}

\author{V.\,B.\,Zabolotnyy}
\affiliation{Institute for Solid State Research, IFW-Dresden, P. O. Box 270116, D-01171 Dresden, Germany}
\author{E.\,Carleschi}
\affiliation{Department of Physics, University of Johannesburg, P. O. Box 524, Auckland Park 2006, South Africa}
\author{T.\,Kim\footnote{Present address: Diamond Light Source Ltd., Didcot, Oxfordshire, OX11 0DE, United Kingdom}}
\affiliation{Institute for Solid State Research, IFW-Dresden, P. O. Box 270116, D-01171 Dresden, Germany}
\author{A.\,A.\,Kordyuk}
\affiliation{Institute for Solid State Research, IFW-Dresden, P. O. Box 270116, D-01171 Dresden, Germany}
\affiliation{Institute of Metal Physics of National Academy of Sciences of Ukraine, 03142 Kyiv, Ukraine}
\author{J.\,Trinckauf}
\author{J.\,Geck}
\affiliation{Institute for Solid State Research, IFW-Dresden, P. O. Box 270116, D-01171 Dresden, Germany}
\author{D.\,Evtushinsky}
\affiliation{Institute for Solid State Research, IFW-Dresden, P. O. Box 270116, D-01171 Dresden, Germany}
\author{B.\,P.\,Doyle}
\affiliation{Department of Physics, University of Johannesburg, P. O. Box 524, Auckland Park 2006, South Africa}

\author{R.\,Fittipaldi}
\affiliation{CNR-SPIN, and Dipartimento di Fisica ``E. R. Caianiello'', Universit\`{a} di Salerno, I-84084 Fisciano (Salerno) Italy}
\author{M.\,Cuoco}
\affiliation{CNR-SPIN, and Dipartimento di Fisica ``E. R. Caianiello'', Universit\`{a} di Salerno, I-84084 Fisciano (Salerno) Italy}
\author{A.\,Vecchione}
\affiliation{CNR-SPIN, and Dipartimento di Fisica ``E. R. Caianiello'', Universit\`{a} di Salerno, I-84084 Fisciano (Salerno) Italy}

\author{B.\,B\"{u}chner}
\affiliation{Institute for Solid State Research, IFW-Dresden, P. O. Box 270116, D-01171 Dresden, Germany}
\author{S.\,V.\,Borisenko}
\affiliation{Institute for Solid State Research, IFW-Dresden, P. O. Box 270116, D-01171 Dresden, Germany}

\begin{abstract}
We present an angle resolved photoemission study of the surface and bulk electronic structure of the single layer ruthenate {\sample}. As the early studies of its electronic structure by photoemission and  scanning tunneling microscopy
were confronted with a problem of surface reconstruction surface aging was previously proposed as a possible remedy to access the bulk states. Here we suggest an alternative way by demonstrating that, in the case of {\sample}, circularly polarised light can be used to disentangle the signals from the bulk and surface layers, thus opening the possibility of investigating many-body interactions both in bulk and surface bands. The proposed procedure results in improved momentum resolution, which enabled us to detect an unexpected splitting of the surface $\beta$ band.  We propose that spin--orbit splitting might be responsible for this, and discuss possible relations of the newly observed surface feature to topological matter.

\end{abstract}

\preprint{\textit{xxx}}
\maketitle

Strontium ruthenates belong to the so-called Ruddlesden--Popper series of layered perovskites\,\cite{Ruddlesden54} and are well-known for their unconventional p-type superconductivity\,\cite{Maeno532},  metamagnetism\,\cite{Perry2661}, proximity to a quantum critical point\,\cite{Grigera11122004, Grigera10122001}  along with the notable effects of spin--orbit coupling\,\cite{haverkort026406, pavarini035115, Oguchi044702,Iwasawa226406}. In particular,
understanding the superconductivity in single layered {\sample}---the first unconventional copper-free oxide
superconductor\cite{Maeno532}---requires a detailed knowledge of its electronic structure.
Active studies by means of photoemission\cite{ Shen180502,Damascelli5194, Iwasawa104514,
Shen187001,Sekiyama060506}, band structure calculations\cite{haverkort026406, Oguchi1385, Singh1358, pavarini035115, Hase3957}, Compton scattering\,\cite{Hiraoka100501} and de Haas-van Alphen measurements \cite{Bergemann639, Mackenzie3786} reached a consensus as regards its low energy
electronic structure: the Fermi surface (FS) of {\sample} consists of three sheets, with the $\alpha$- and $\beta$- sheets formed by quasi-one-dimensional (1D) out-of-plane Ru\,4d$_{yz}$ and 4d$_{zx}$ orbitals, whereas the $\gamma$ sheet is formed by  the two-dimensional (2D) in-plane Ru\,4d$_{xy}$ orbitals.

A characteristic feature of {\sample}, which initially was quite perplexing to the photoemission community, is a $\sqrt{2}\!\times\!\!\sqrt{2}$ reconstruction due to slight rotations of the RuO$_6$ octahedrons in the  topmost layer\cite{Matzdorf746}. The reconstruction  implies  doubling of the unit cell and thus folding of the surface Brillouin Zone (BZ). As a result, a new set of surface-induced states with different underlying dispersions appears, so that both signals are seen superimposed in a typical ARPES experiment\cite{Ingle205114}. No other Fermi surfaces, either bulk- or surface-related have been identified in {\sample} until now.

To overcome the problem of surface related states it was suggested to cleave the sample at high temperature ($\sim$200K), or age the sample surface after a low-temperature cleave. This  recipe have been followed  by the majority  of  the ARPES community \cite{Iwasawa200673, Iwasawa104514,  Shen180502, Shen187001, Damascelli5194, Aiura117005, Kidd107003}. However, in the most recent STM study\,\cite{pennec216103} it was argued that high temperature cleaving does not actually suppress  the $\sqrt{2}\!\times\!\!\sqrt{2}$ surface reconstruction. Instead it was suggested that the major aging effect is due to the increased surface disorder on the mesoscopic scale that effectively blurs the superstructure replicas, so that they become less visible in  ARPES Fermi surface (FS) intensity maps. Obviously such surface disorder equally  scatters photoelectrons not only from the replicas, but also from the original bulk bands, resulting in a disorder-induced broadening of photoemission peaks \cite{Shevchik3428, Theilmann3632, Cerrina1798}.
 To account for these adverse effects, we have performed measurements at extremely low temperatures ($T<2$\,K),
analysing spectra measured from both `aged' and `fresh' samples.
We find that instead of the earlier proposed remedy of high temperature cleaving, one may rely
on use of circularly polarized light  to establish the  origin of bulk and surface features. Owing to the minimized surface degradation we  now observe bulk $\alpha$, $\beta$, $\gamma$ bands and their surface counterparts along with
an additional new  feature. According to its dichroic pattern, the new feature must be yet another surface counterpart of the $\beta$ band. Since there are numerous examples where the surface state undergoes splitting due to the spin--orbit
interaction we suggest that fully relativistic calculations might be needed to understand the origin of the new state.

\section{Methods}
\begin{figure}[t]
\includegraphics[width=1\columnwidth]{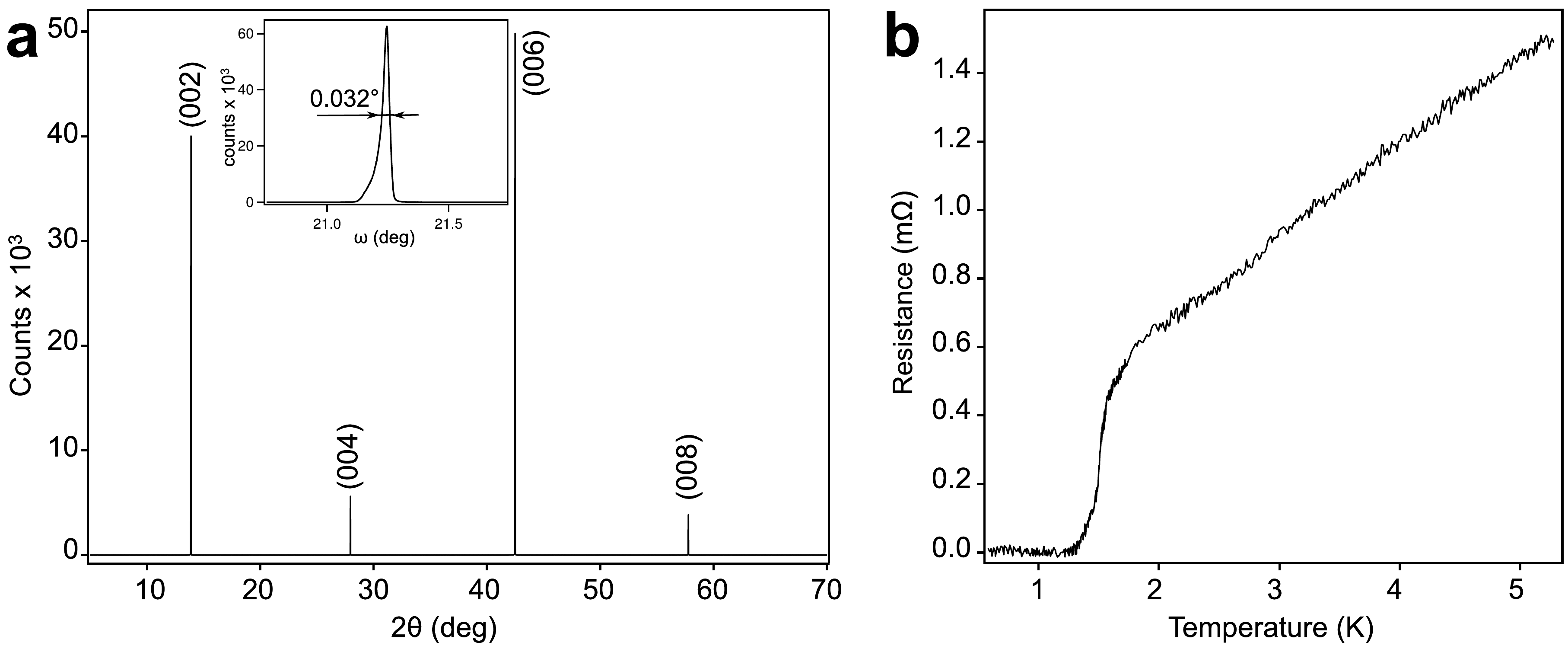}\\
\caption{$\!\!$. (a) $\theta$--$2\theta$ XRD pattern on the (001) surface of {\sample}. The inset contains the 2$\theta$--$\omega$ rocking curve scan. (b) Resistance of  {\sample} samples as a
function of temperature.
}
\label{x_ray}
\end{figure}

The composition of the samples used in this study has been characterised by X-ray diffraction and electron backscatter diffraction. The structure and crystalline qualities were assessed by a high-resolution X-ray diffractometer (Philips, model X$'$Pert MRD), with Cu K-$\alpha$ source. The typical XRD pattern taken on a cleaved surface of the
{\sample} crystals is shown in Fig.\ref{x_ray}a.  All the diffraction peaks can be identified with the expected (001) Bragg reflections of the {\sample} phase, confirming the absence of any spurious phase. Moreover, the high quality of the crystals is also confirmed by the narrow peak width in the rocking curve shown in the inset to Fig.\ref{x_ray}a (FWHM = 0.032$^\circ$). The purity of the crystals is supported by a.c. susceptibility and resistivity measurements (Fig.\ref{x_ray}b), where the narrow superconducting transition with
$T_\textup{c}$=1.34\,K, is a signature of a low impurity concentration \cite{Kikugawa237}.

All photoemission measurements were performed at the BESSY 1$^3$ ARPES station equipped with a SCIENTA R4000 analyzer and a Janis $^3$He cryostat. Spectra presented in this manuscript were recorded from high quality {\sample}  samples cleaved at high/low temperature. For the high temperature cleave the samples were cleaved on the transfer  arm at  $T$=300\,K, just before transferring them to the cold finger of the cryostat. For the low temperature cleave the samples were first mounted on the cold finger of the cryostat and after pre-cooling down to $T\sim15$--$40$\,K the cleave was performed.
Sample orientation and determination of high symmetry directions were done using wide overview Fermi surface maps, one of which is shown in Fig.\,\ref{overview}. This method allows for accurate determination of the initial offset angles of the sample manipulator.

\section{RESULTS AND DISCUSSION}
\begin{figure}[b]
\includegraphics[width=0.999\columnwidth]{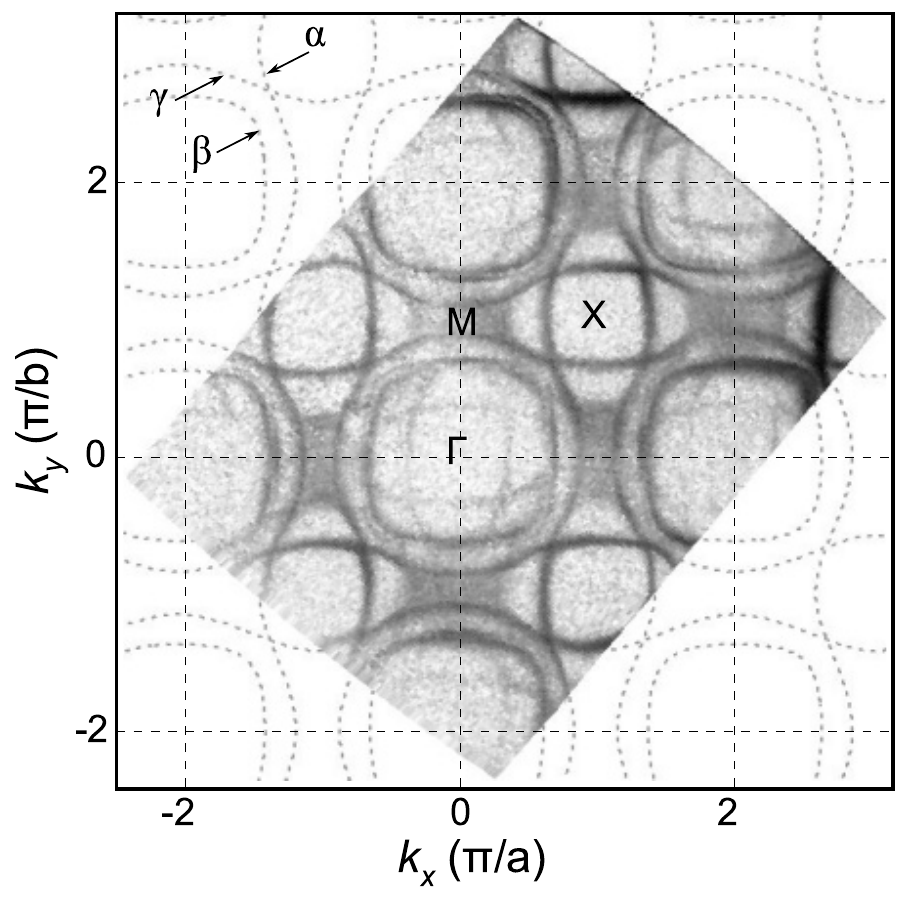}\\
\caption{$\!\!$. Typical overview Fermi surface map used for sample orientation. $\alpha$, $\beta$ and $\gamma$ denote three bands contributing to the Fermi surface of {\sample}. The underlying set of contours extend the experimentally observed bands over the whole extended Brillouin zone picture. The map was measured using horizontally polarised light with $h\nu$ = 100\,eV.}
\label{overview}
\end{figure}

In Fig.\,\ref{overview}  we present the experimental FS map given by the distribution of the photoemission intensity at the Fermi level (FL) recorded over a voluminous part of the reciprocal space in the superconducting state of {\sample}. The dark features correspond to regions where the bands cross the FL. In agreement with the earlier measurements, one can identify the square-like contour centered at the $\mathrm{X}$ point as corresponding to the $\alpha$ band.  The other two, more rounded features centered at the  $\Gamma$ point, must be formed by the $\beta$- and $\gamma$- bands. The surface reconstruction results in the appearance of replica bands, shifted by the vector $\overline{\Gamma\mathrm{X}}$.

\begin{figure}[t]
\includegraphics[width=1\columnwidth]{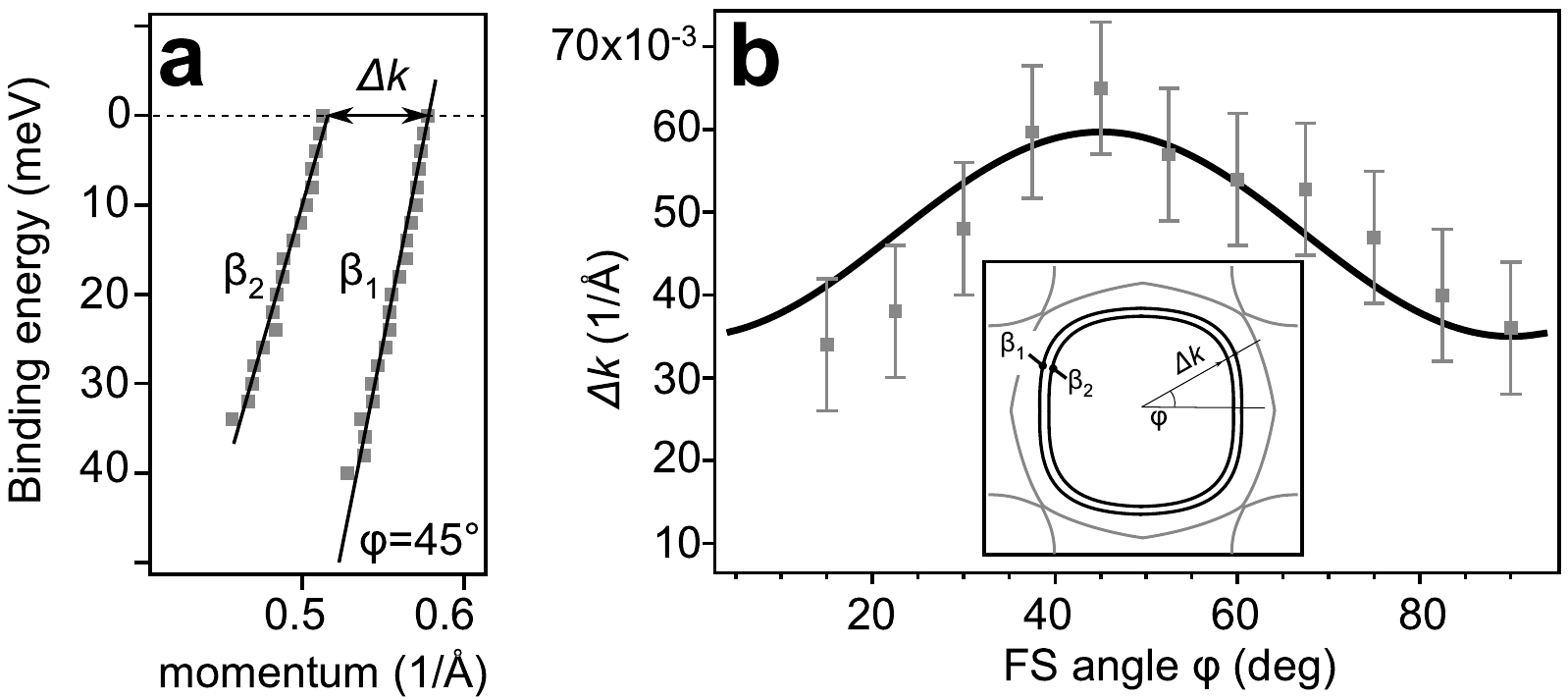}\\
\caption{$\!$.  Momentum splitting between the $\beta_1$ and $\beta_2$ bands along the Fermi surface contour. (a) Extracted MDC dispersion for the energy--momentum cut at $\varphi=45^\circ$ and momentum splitting at the FL. The square symbols are the MDC peak positions for the two features. (b) Momentum splitting at the Fermi level, $\Delta k$, over the Fermi contour. The black line is the fit to the function $A$cos($\varphi$)+$B$.}
\label{splitting}
\end{figure}

  A closer look reveals that the picture is more complex. The first and the most obvious detail can already be  seen in the FS map (Fig. \ref{overview}).  The FS contour corresponding to the $\beta$ band appears to be split, with the splitting most notable along the diagonal of the BZ. The value of the momentum splitting between the two features can be followed in Fig.\,\ref{splitting}.

\begin{figure*}[!t]
\includegraphics[width=1.5\columnwidth]{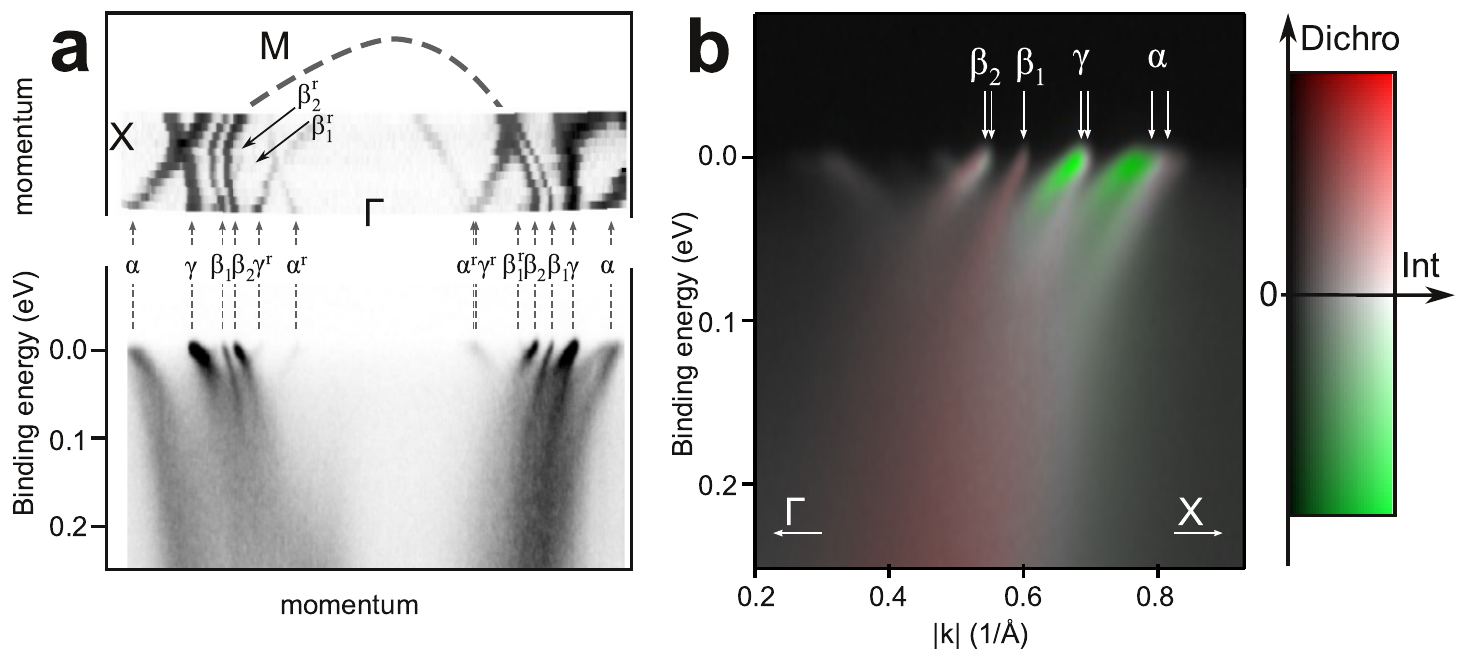}\\
\caption{(color). (a) Fermi surface map and energy--momentum image, used to index features seen in spectra, $h\nu=50$\,eV, $T=1.3$\,K (b) Circular dichroism. The intensity in the false color image  represents the sum of the signals obtained with circular right (CR) and circular left (CL) polarised light, while the  colour corresponds to the dichroic signal CR$ - $CL.  }
\label{dichro}
\end{figure*}

In Fig. \ref{dichro}a we show the FS measured with lower excitation energy, to get better effective energy and momentum resolution, using a sample cleaved at low temperature $T=16$\,K. The FS map is supplemented with an energy--momentum cut, which  allows one to trace the energy dispersion of the spectral features. To classify all the observed bands,  we label the trivial $\alpha$ band replicas arising due to the surface reconstruction as $\alpha^\textup{r}$, the two features apparently related to the $\beta$ band are denoted as $\beta_1$ and $\beta_2$, while the features related to the $\alpha$ and $\gamma$ are labeled accordingly.
Considering the energy--momentum cut it is easy to see that also the $\alpha$ and $\gamma$ features, that appear as
individual bands in the FS map, actually consist of pairs of bands with slightly different Fermi velocities.

Obviously, the multitude of the features we observe in the spectra must result from a superposition of bulk and surface states. A very effective method to discern between them is based on the use of  circularly polarised light. It was shown, both experimentally and using simple theoretical considerations, that the nontrivial pattern of the electromagnetic field at the solid--vacuum interface\cite{ Pforte115405}  leads to a notable circular dichroism for  states primarily localised at the  surface, with negligible effect on the bulk states\cite{zabolotnyy024502}. Previously  we have successfully applied this method to distinguish between the bulk and surface contributions in another layered superconductor YBa$_2$Cu$_3$O$_{7-\delta}$. In particular, it was possible to disentangle the spectral features corresponding to the topmost overdoped layer of YBa$_2$Cu$_3$O$_{7-\delta}$ from the bulk bands undergoing the superconducting transition\cite{zabolotnyy064519, Zabolotnyy2007888}. Here we utilise the same method to separate the surface and bulk bands in {\sample}.

The results are shown in Fig.\,\ref{dichro}b. In order to facilitate the comparison between the bands exhibiting circular dichroism and those with negligible dichroism,  we plot them in one image, where the brightness corresponds to the sum of intensities obtained with opposite polarisations (CR$+$CL) and the colour, ranging from green through white to red, encodes the dichroism strength (CR$-$CL). As can be seen from the map shown in Fig.\,\ref{dichro}a, the energy--momentum cut we have selected is optimal for highlighting all features. The rightmost pair of dispersing features forms the $\alpha$ pocket of the FS.  Now with the dichroic pattern we can see that the two bands are qualitatively different. The steeper band exhibits virtually no dichroism, based on which we conclude that this must be a bulk band. The slowly dispersing band is strongly dichroic, therefore we believe this one has to be a surface related counterpart of the $\alpha$ band. This identification  is in agreement with earlier experimental work where the double structure of the $\alpha$ band was resolved \cite{Ingle205114} as well as with the theoretical calculation from Ref. \onlinecite{Shen180502}.

Exactly the same dichroic pattern is also observed for the $\gamma$ feature, except for the fact that the splitting at the
Fermi level is negligibly small, so that owing  to the difference in the Fermi velocities the two counterparts can only be seen clearly separated at binding energies close to 50\,meV.  Here again the surface component has a higher renormalisation as compared to that of the bulk.

\begin{figure}[b]
\includegraphics[width=1\columnwidth]{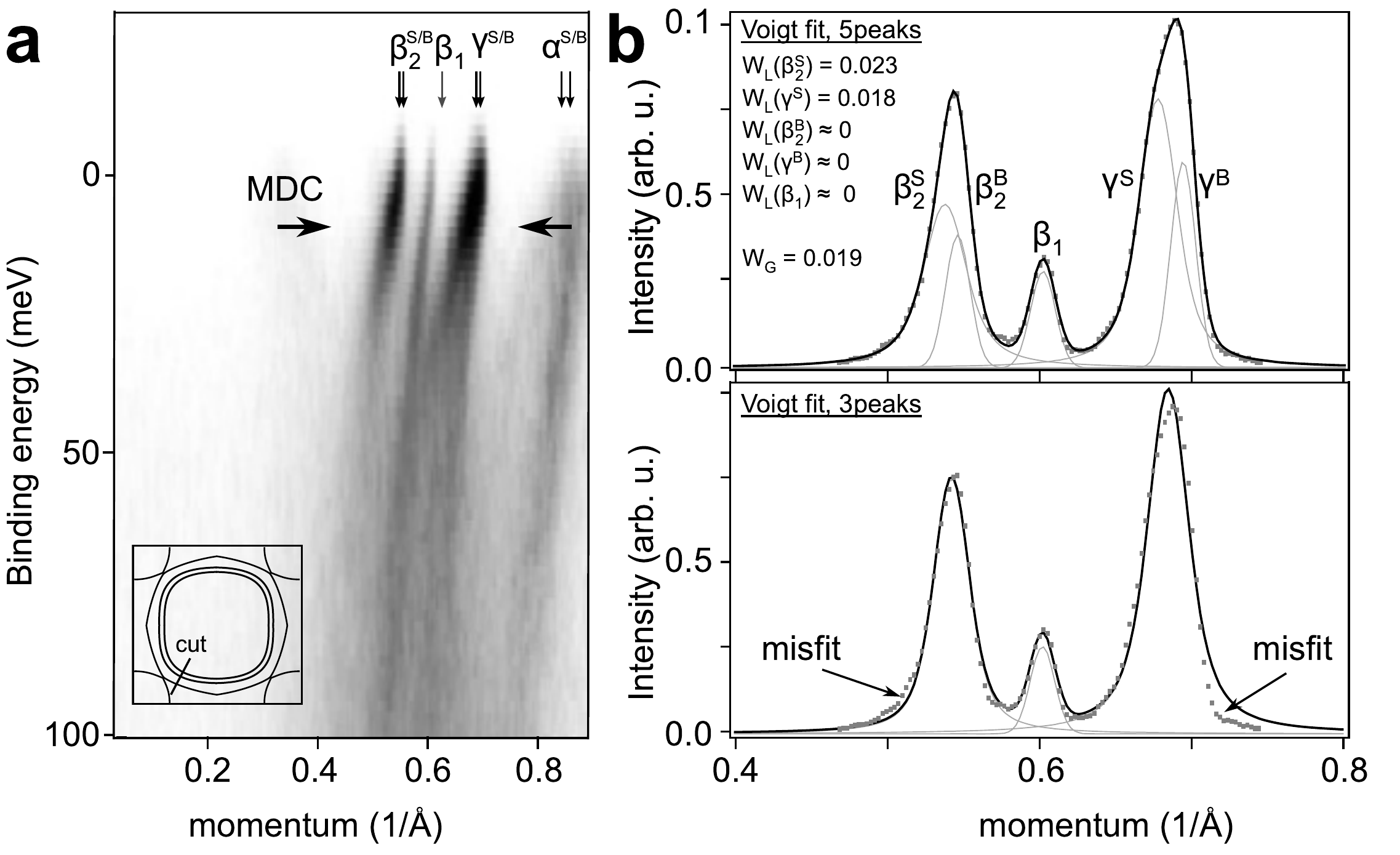}\\
\caption{$\!\!$. (a) Energy--momentum cut through the $\beta_1$, $\beta_2^{S/B}$ and $\gamma^{S/B}$ bands. The arrows denote the energy position and momentum range for the MDC shown in (b). The symbols represent the experimental data, while the fit with the five and three Voigt profiles is shown by the line. The Gaussian FWHM for the five-peak fit was held constant at the value corresponding to the experimental angular resolution, while all other parameters were left free. }
\label{MDC_fits}
\end{figure}

According to the dichroic image, the feature corresponding to the $\beta_2$ band consists of two
components as well, but unlike the $\alpha$  band the splitting between them is quite small, approaching approximately 0.015 \AA$^{-1}$ at the Fermi level.  In the dichroic pattern (Fig.\,\ref{dichro}b) this is manifested as a red shade on the left side, and a white shade on the right side of the composite surface+bulk $\beta_2$ feature.
Despite its small size, the splitting of the $\beta_2$ feature can also be infered if one compares the width of the $\beta_2$ feature to the width of the single $\beta_1$, as the former appears notably broader (see FS map in Fig.\,\ref{dichro}a). This argument can receive stricter development in the form of line shape analysis. In Fig.\,\ref{MDC_fits} we show an MDC containing contributions from $\beta_1$, $\beta_2$ and $\gamma$ features. Assuming that both $\beta_2$ and  $\gamma$ features are split into surface and bulk counterparts, the MDC shown in Fig.\,\ref{MDC_fits} can be nicely fitted by 5 Voigt profiles\cite{Evtushinsky172509}. At the same time the 3 peak model, i.e. no splitting of the $\beta_2$ and  $\gamma$ features,  results in obvious misfits for both $\beta$ and  $\gamma$ features, suggesting their splitting, in agreement with the conclusion drawn from the dichroic data.

Therefore we see that the composite surface+bulk structure holds also for the $\gamma$- and $\beta_2$-pairs with progressively smaller splitting. In particular this gradual decrease in momentum splitting  can be seen in the notably broader Fermi surface contour for the composite $\alpha$ pocket as compared to the other bands (see Fig.\,\ref{overview} and Fig.\,\ref{dichro}a.).  All three surface components are also seen replicated in the new Brillouin zone. When going from the $\Gamma$- to the  $\textup{X}$-point in Fig.\,\ref{dichro}a,  first the  $\alpha$- and $\gamma$- replicas are seen to cross, then comes the the $\beta_2^\textup{r}$-replica which forms a tiny lens when considered together with the barely split $\beta_2$-pair. This is in agreement with earlier ARPES studies\cite{Damascelli5194, Shen180502, Shen187001}.
In this light the feature that we labeled as $\beta_1$ appear to be special as it does not fit into  the simple picture of three main bands and their replicas, whereas the strong dichroism (red color)  points to its surface origin.

At present we cannot unambiguously determine the origin of this feature. Owing to the strength of the dichroic pattern this must be yet another surface counterpart of the bulk $\beta$ band. Also the  narrower momentum width suggests a higher quasiparticle life-time and negligible $k_z$ dispersion, which both would be distinct properties specific to a true surface state. As mentioned above, this would also imply that the $\beta$ surface band undergoes unexpected splitting, in contrast to the $\alpha$ and $\gamma$ surface bands. To check if this is plausible in the first place, it is useful to recall that the spin--orbit interaction plays an important role in determining the dispersion of low energy bands in {\sample} \cite{Oguchi044702, pavarini035115, haverkort026406, Liu_026408, Ng49}.  It is also worthwhile to note that spin--orbit coupling can induce splitting of a surface state;  Au(111) \cite{Nicolay033407},  Li/W(110) and  Li/Mo(110) \cite{Rotenberg4066} being good examples, where the surface state undergoes  splitting similar in magnitude  to that between the surface $\beta_2$ and $\beta_1$ features.  Moreover, the case of Bi(111) \cite{Koroteev046403} demonstrates that the spin--orbit induced  splitting can even be much stronger than the one  currently observed.

Owing to its apparent surface origin it is tempting to discuss the newly observed surface state $\beta_1$, in the light of topological matter that is currently in  focus in the solid state community \cite{Nobukane144502, Jang331, Wray855, Wray32, Hasan3045}. Purely phenomenologically,  we notice that the new $\beta_1$ feature carries a set of typical spectroscopic signatures characteristic of topological states\cite{Wray855, Wray32, Hasan3045}. Namely, it forms a single electron-like FS around the $\Gamma$-point, it is robust with respect to
different ways of surface preparation (cleavage temperature), and  it has narrow momentum width and linear dispersion in the vicinity of the FL\cite{topology}.  Therefore, along with at least two possible ways to realize a  topologically non-trivial surface states in {\sample}, one may expect the new state to be connected to topological superconductivity. The first possibility is that in the chiral superconductor with broken time reversal (TR) symmetry, the gapless edge states can emerge as in a Quantum Hall Effect (QHE)\cite{Gurarie, Read}.  Another one allows for restoration of the TR symmetry at the surface\cite{Tada} and two gapless counter-propagating modes carrying the spin-current in $k_x\pm ik_y$ type TR invariant superconductors with odd parity\cite{Schnyder, Fu_Berg, Tada}. However this would also require the $\beta_1$ feature to be located in the superconducting gap, which is not the case. Instead, we see that the discovered state can be  hosted by the so-called correlation gap, as the experimental bandwidth of the conduction band has been shown to be strongly renormalised by at least a factor of two\cite{pchelkina035122, Sekiyama, Yokoya13311},  resulting in a band gap between approximately 0.8\,eV and 3\,eV. Thus the $\beta_1$ feature may actually be the upper part of the Dirac cone situated in the band gap of the order of 2\,eV with its Kramers point at approximately 0.75\,eV binding energy. The Kramers point, however, is not distinctly observed because of the proximity of the $\beta$ and $\gamma$ bands, which is similar to  Bi$_2$Se$_3$, where bulk states at the FL cause broadening of surface states, and progressive electron doping of the topological insulator results in their gradual blurring\cite{Wray32}.
This implies that  hypothetical hole doping may shift the the chemical potential into the correlation gap and bring  \sample into the regime of a topological insulator. Therefore we note that a spin resolved ARPES study able to determine
the spin texture of the $\beta_1$ feature\cite{Hsieh919}  together with a careful computational approach, which takes into account the strong correlations and spin--orbit interaction may  put a solid ground under the suggested here surmises.

\begin{figure}[t]
\includegraphics[width=0.75\columnwidth]{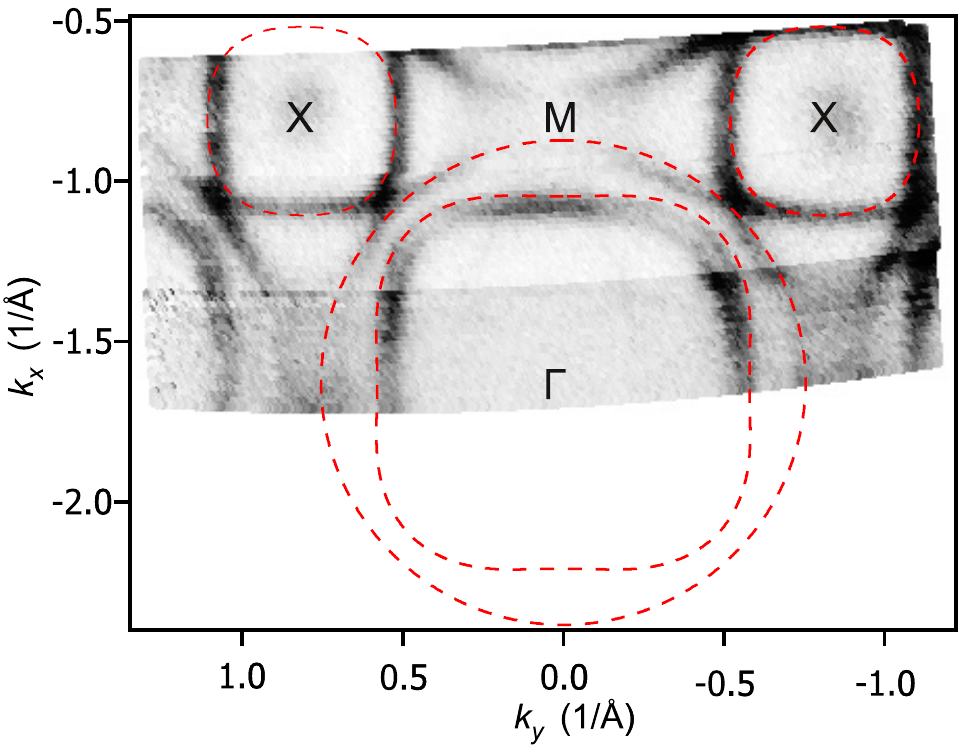}\\
\caption{$\!\!$. Fermi surface map measured at 40\,K from the `aged' sample cleaved at 300\,K. Note the vanishing of the $\alpha$ band replicas. The overlapped contours are the results of de Haas-van Alphen fits \cite{Bergemann639, Mackenzie3786}.  }
\label{doping}
\end{figure}

Since the experimental Fermi surface of {\sample}  is measured in more detail now,  it is instructive to see how our observation relates to the FS topology that was reconstructed from bulk sensitive de Haas-van Alphen measurements \cite{Bergemann639, Mackenzie3786}. In Fig.\,\ref{doping} we show ARPES FS acquired from the sample cleaved at 300K with the de Haas-van Alphen contours overlapped over them.

One interesting observation is that the form of the $\alpha$ pockets appear more square-like, as compared to the
rounded form suggested by  the  de Haas-van Alphen fits.  This  particularity of the $\alpha$ band dispersion may turn out to be important when making detailed comparisons to theoretical results, since the BZ diagonal is the direction where one expects the largest corrections due to spin--orbit interaction \cite{Oguchi044702, pavarini035115, haverkort026406}.

Another interesting observation is that the new $\beta_1$ feature is actually closer to the de Haas-van Alphen data. This contrary to the previous discussion would suggest that the  $\beta_1$ feature has to be the true $\beta$ bulk band.
This would also imply  different identification of what we call a $\beta_2^{\tiny{\textup{B}}}$  bulk band in the pair of $\beta_2$ features. One may argue that both $\beta_1$ and $\beta_2^{\tiny{\textup{B}}}$ originate from different $k_z$ cross sections of the 3D sheet of the $\beta$ band, but the distance between the $\beta_1$ and  $\beta_2^{\tiny{\textup{B}}}$ is too large to be explained by the $k_z$ dispersion as measured by de Haas-van Alphen.
Nonetheless the identification of the pair of  $\beta$ features as purely surface derived does not solve all the controversies. While the high temperature cleave seems to completely suppress the surface related replicas of the $\alpha$ band (see Fig.\,\ref{doping}) both  $\beta$ features remain equally well visible. Therefore we believe that a complete calculation, which includes both surface effects and the effects of spin--orbit coupling would be necessary to find the true nature of the new $\beta_1$ feature.

\section{Summary \& Conclusions}
Despite common opinion on the extreme surface sensitivity of angle resolved spectroscopy, and the detrimental effects of surface states introduced upon sample cleavage, we have demonstrated that the method can be successfully tuned to study bulk as well as surface states. In a certain sense, the suggested strategy  outbalances the commonly used surface degradation approach, preserving the high energy and momentum resolution of the method.  This allowed us to detect a new feature in the electronic structure of {\sample}: the unexpected surface band, which is likely to be related to the strong spin--orbit coupling effects in this compound. The idea of a possible link of the new surface state to topological matter was discussed.

\section{Acknowledgements}
We acknowledge useful discussions with Prof.\,Jeroen\,van\,den\,Brink,   Dr.\,Klaus\,Koepernik and Dr. Felix Bauberger. The pro\-ject was supported in part by grant ZA 654/1-1.   E.\,C. and B.\,P.\,D. thank the Faculty of Science at the University of Johannesburg for travel funding. M.\,C., R.\,F. and A.\,V. acknowledge support and funding from the FP7/2007-2013 under grant agreement N.264098-MAMA. V.\,B.\,Z. acknowledges EU support during measurements at the synchrotron facility Helmholtz-Zentrum Berlin f\"{u}r Materialien und Energie HZB.

%\bibliographystyle{PRBstyle}
%\bibliography{state}

\end{document}